\documentstyle[epsfig,epsfig]{aipproc2}
\begin{document}

%%%% SPECIAL MACROS

\def\la{\mathrel{\hbox{\rlap{\hbox{\lower4pt\hbox{$\sim$}}}\hbox{$<$}}}}
\def\ga{\mathrel{\hbox{\rlap{\hbox{\lower4pt\hbox{$\sim$}}}\hbox{$>$}}}}

\def\ha{H$\alpha$}
\def\msunyr{M$_{\odot}$ yr$^{-1}$}

{ \hyphenpenalty=5000 \noindent To be published in {\it The
Ultraviolet Universe at Low and High Redshift}, ed. W. H. Waller
(AIP Press), 1997.} 

\vspace{-0.2in}

\title{THE ULTRAVIOLET MORPHOLOGY OF GALAXIES}

\author{Robert W. O'Connell   }
\address{Astronomy Department, University of Virginia\\ Charlottesville,
VA\ \  22903-0818  }

%\lefthead{LEFT head}
%\rig436thead{RIGHT head}

\maketitle

\vspace{-0.15in}

\begin{abstract} The vacuum ultraviolet offers a unique perspective on
galaxy morphology, stellar populations, and interstellar material
which is of particular relevance to interpreting high redshift
galaxies and the history of cosmic star formation.  Here we review UV
imaging studies of galaxies since 1990.  

\end{abstract}

\section*{Introduction}

The concepts of morphology and classification are closely linked.
Classification is one of the most powerful organizing tools ever
developed by science, but as it is applied to galaxies, morphology is
more than simple classification.  It is a means to inferring the
evolutionary state and history of galaxies.  There is a set of
conventional associations we have learned to make between morphology
and the inferred nature of galaxies.  Regularity or symmetry are taken
to imply stability, i.e.\ some kind of equilibrium state.  There is,
for instance, the powerful notion that in such situations we can infer
the three-dimensional structure of galaxies from their two-dimensional
appearance.  A de Vaucouleurs distribution of surface brightness [$I
\sim \exp(-r^{0.25})$] is taken to imply a spheroidal mass
distribution, whereas an exponential profile [$I \sim
\exp(-\alpha r)$] is taken to imply a disklike mass distribution.  
On the other hand, irregularity or asymmetry is assumed to imply a
non-equilibrium state or a low-mass system in which random velocities
are significant compared to the organized velocity field.  

I have deliberately used tentative language in describing these
associations because essentially all of our intuition about them has
been developed from observations in a narrow range of visible
wavelengths (mainly 4500--6500 \AA).  There is an implicit assumption
that V-band light is a good tracer of the dynamical mass of galaxies.
It is now widely acknowledged that this cannot be strictly correct
because of the presence of significant amounts of dark matter.  It is
less well appreciated that even the luminous, stellar component of
galaxies is only well traced by V-band light if $M/L_V \sim const$,
which, again, cannot be strictly true.  One of the main motivations for
studying the morphology of galaxies in bands other than the visual is
to provide an independent perspective on the interpretation of V-band
morphologies---to test the assumptions and provide new insights.  The
availability of multiband data for galaxies can make up to some degree
for the fact that astronomical morphology is limited to two
dimensions.  In recent years, it has been possible to assemble a small
but representative sample of images of nearby galaxies in the vacuum
ultraviolet.  In this paper, I review what we know
about UV galaxy morphology, based mainly on the images returned by the
{\it Astro} Ultraviolet Imaging Telescope.

\section*{The Ultraviolet Viewpoint}

In what ways can we expect the ultraviolet to yield new information on
galaxy structures and histories?  Briefly, the vacuum UV (912--3200
\AA) provides {independent and unique} information on the character of
hot star populations, dust, very high or very low temperature gas, and
AGN's.  To elaborate: 

\begin{itemize}

\item The UV allows {\it direct detection} of the massive stars responsible
for most ionization, photodissociation, kinetic energy input, and
element synthesis in galaxies.

\item The UV has the highest sensitivity of any spectral region to
stellar temperature and metal abundance, implying that it is uniquely
valuable to characterize stellar populations, star formation rates
(SFR's), and SF histories.  Figure 1 illustrates the strong evolution
of UV energy distributions of stellar populations over timescales up
to 3000 Myr.  The UV is a sensitive indicator of SF histories over the
past $\sim$5000 Myr.  Its usefulness extends even to ``old'',
quiescent systems such as elliptical  galaxies
\cite{dor97}.  By contrast, the optical emission lines which are
widely used to infer SFR's reflect activity only over the past $\sim
5$ Myr, after which photoionization rapidly decreases.  The infrared
continuum ($\lambda > 1\mu$) is primarily an indirect measure of
SFR's, involving either downconversion of absorbed photons (not
necessarily from young objects) by dust grains or complex
supergiant evolution.  AGN's and cold interstellar cirrus have both
acted to confuse IR estimates of SFR's.

\begin{figure}[t!] % fig 1
\centerline{\epsfig{file=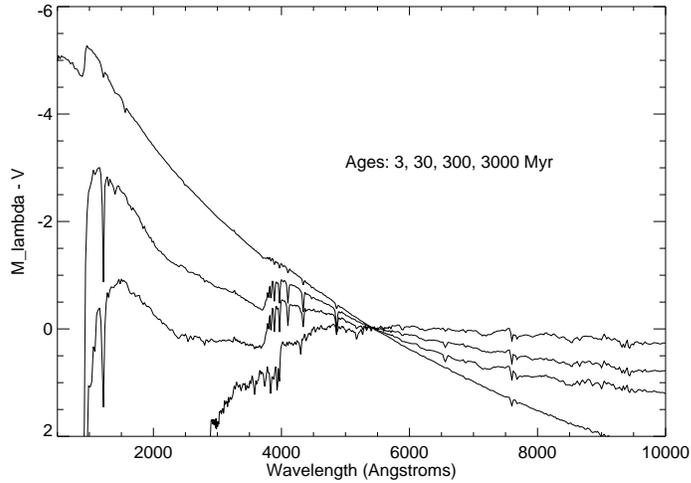,height=2.7in,width=3.8in}}
\caption{ Synthetic spectral energy distributions of single generation
populations for ages 3--3000 Myr from [1], showing the rapid 
evolution in UV amplitude and shape. Note that the shape of the 
near-IR spectrum ($\lambda > 7000$) is much less sensitive to age.  
}
\vspace*{2pt}
\label{fig1}
\end{figure}

\item The UV offers high sensitivity to interstellar dust and regions
of concentrated cold material (i.e.\ potential star-formation sites).
Before the advent of UV observations of galaxies, it was widely
assumed that this would actually be a serious disadvantage for the UV
because the Galactic extinction law yields $A(UV)/A(V) > 2.5$; some
expected disk galaxies to be dark in the UV.  However, it turns out
(see below) that dust is not dominant in the UV morphology of galaxies
except where the dust layers are disturbed.  The UV may ultimately
prove to be a valuable tracer of quiescent, cold,
molecular material.  Conventional tracers, like radio-emitting CO,
occur only where there is significant dissociation and, as
emphasized by
\cite{all97} and others, may be unreliable indicators of total H$_2$
mass.  Because the albedo of dust is high in the UV, dense, cool
interstellar clouds far from star forming regions can be detected by
scattered light, as in the case of the faint gaseous arms of M101
\cite{don81,stech82}.  UV imaging can also detect H$_2$ in
photodissociation regions directly by virtue of its fluorescence bands
(1550--1650 \AA)  \cite{witt89,mart90}.

\item  The UV contains uniquely important emission line probes
of interstellar gas in the  T $\sim$ 10$^5$-10$^6$ K regime,
including C IV ($\lambda$1550), N V ($\lambda$1241), and O VI
($\lambda$1035).  These have not been exploited much yet
in galaxy imagery, though \cite{corn92} and \cite{hill95}
discuss C IV images of supernova remnants.  
   
\item The UV provides excellent isolation of hot sources from the
often overwhelming optical/IR cool star background of galaxies.  This
offers greatly improved detectability for minority populations of
massive stars (e.g.\ in elliptical galaxies); hot horizontal branch
and related late stages of stellar evolution such as planetary
nebulae; low luminosity AGN's (e.g.\ \cite{maoz96}); star clusters;
nonthermal jets; and other hot sources.  

\item A minimum in the natural night sky background occurs at
1600-2400 \AA.  This is the deepest window in the UV-optical-IR
spectrum and permits detection of extremely low surface brightness
objects, perhaps up to 100,000$\times$ fainter than the ground-based
sky \cite{oc87}.  

\end{itemize}

\section*{Applications at High Redshifts}

For two reasons, some of the most important applications of UV imaging
of nearby galaxies will be to galaxies at high redshift.  First, the
restframe UV continuum is the best tracer of star formation and is
measurable to very high redshifts (z $\ga$ 10) with optical/IR
instruments.  For instance, the 2800~\AA\ restframe continuum has been
used to estimate the cosmic star formation density at $z \sim$ 0.5--2
for the CFRS, Hubble Deep Field, and other surveys
\cite{pei95,lil95,mad96}, leading to the conclusion that most gas
processing occurred at relatively recent epochs, $z \sim 1$--2.
Second, observations of high-$z$ galaxies are preferentially made in
the restframe UV.  This is particularly true for ground-based
telescopes, where the rapidly increasing night sky brightness seriously
compromises observations in the near infrared ($\lambda > 7000$\AA).
Because galaxy appearance is a strong function of wavelength (as
illustrated below), this means there is a large ``morphological
k-correction'' which must be calibrated in assessing the morphologies
of distant objects.  To what extent is the strange appearance of many
distant galaxies the product of simple bandshifting?  Preliminary
explorations suggest that distant field galaxies are in genuinely different
states of morphological evolution \cite{boh91,ab96,gia96}.

To take advantage of this remarkable opportunity to study early galaxy
evolution at $z > 1$, we require comprehensive fiducial UV
studies of nearby galaxies to assess the astrophysical drivers of UV
luminosity, the cosmic star formation history over the past few Gyr,
the morphological k-correction, surface brightness selection effects
(a very serious problem at high redshifts since $I
\sim (1+z)^{-4}$), and the effects of reduced spatial resolution on
high-$z$ morphologies.  

\section*{UV Imaging Instruments}

Extragalactic UV astronomy to date has been largely based on
spectroscopy, usually with small entrance apertures (e.g. IUE,
HST/FOS), or low-resolution photometry (e.g. OAO, ANS, FAUST
\cite{deharv94}).  The first UV {\it image} (i.e.\ with many
resolution elements) of another galaxy (the LMC, which exhibits a
remarkable UV/optical morphological change) was obtained by the NRL
Apollo S201 camera from the lunar surface in 1972
\cite{pc81}.  But progress in imaging up to 1990 was relatively slow
(reviewed in
\cite{oc91}).  Since 1990, we have accumulated a sample of vacuum UV
images of about 200 galaxies, principally from three instruments:
{\it (i)} the HST/FOC, which has better red leak rejection than 
HST/WFPC2 and which has recently produced an atlas of the nuclei
($22\arcsec$ fields) of 110 nearby galaxies at 2300 \AA\ with
$0.05\arcsec$ resolution \cite{maoz96}; {\it (ii)} the SCAP/FOCA
balloon-borne telescope of the Marseille and Geneva groups, which
operates in a narrow band near 2000 \AA\ with a field of view of
1.5$^\circ$ and resolution of $15\arcsec$ 
\cite{mill91,bers94}; and {\it (iii)} the Ultraviolet Imaging
Telescope (UIT) of the {\it Astro} Spacelab missions \cite{stech97}.
The images I will present here are all from the UIT, although the
basic conclusions are common to all three sets of data.  

UIT flew on two Spacelab missions in 1990 and 1995 as one of the
three, co-pointed UV telescopes of the {\it Astro} observatory.  It
has a $40\arcmin$ diameter field and $3\arcsec$ resolution in two
solar-blind broad-band channels (the ``FUV'' at 1500 \AA and the
``MUV'' at 2500 \AA).  Its monochromatic UV limiting magnitudes in
30-min exposures are $\sim 18$ for point sources ($V \sim 23$ for hot
stars) and 26--27 mags per square arcsec for extended sources.  Other
information can be found in
\cite{stech97}.  Useful data was obtained for some 80 nearby
galaxies.  We are combining the UV data with ground-based optical CCD
imagery of comparable resolution \cite{cheng96} to produce two {\it
UV/Optical Atlases of Galaxies}, which will include both images and
photometry \cite{marc97}.  The subsequent figures in this article are
taken from the {\it Atlases}.

\section*{UV Morphologies of Nearby Galaxies}

\begin{figure}[t!] % fig 2
\centerline{\epsfig{file=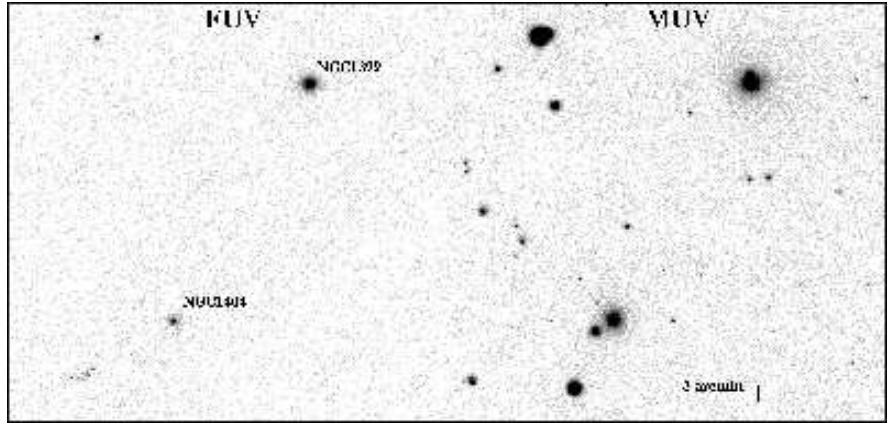,height=2.2in,width=4.6in}}
\vspace*{4pt}
\caption{ 
UIT images of the Fornax cluster ellipticals NGC 1399 and 1404.  The
mid-UV band is dominated by the main sequence turnoff.  All of the
far-UV light is from the UVX component.  It is smooth, without
evidence for massive stars, though is more concentrated than the mid-UV
light.  Note that the foreground stars have mostly vanished in
the FUV; this is a pictorial representation of how rare are the
objects which make up the UVX. }
%%\vspace*{1pt}
\label{fig2}
\vspace{-0.15in}
\end{figure}

\noindent {\bf Elliptical Galaxies}:  One of the first important
discoveries of UV astronomy was that spiral bulges were bright at
far-UV wavelengths where their normal main sequence and giant branch
stars produce negligible flux \cite{cw79}.  This ``UVX'' was later
found in all nearby E/S0 and old bulge populations.  Massive stars are not
the source of the UVX.  The far-UV
brightness profiles (see Fig. 2) are smooth and unresolved and usually
have $r^{0.25}$-type spheroidal profiles \cite{boh85,oc92}. 
Some objects have nearby exponential UV profiles, but the UV isophote
shapes match those in the optical, and a genuine disk is probably not
present \cite{ohl97}.  UV spectra also do not have massive star
signatures.  The best evidence is that the UVX is produced by
extreme horizontal branch stars and their descendants in the low-mass,
metal-rich population which dominates at longer wavelengths
\cite{b3fl,dor95,brown95,yi97}.  The EHB stars are hot by virtue of having
very thin envelopes, and it is a remarkable circumstance, if this
interpretation is correct, that relatively crude UV observations can
determine envelope masses for stars in galaxies many Mpc distant with
a precision of a few $0.01\, {\rm M}_\odot$. 

The UVX varies more between galaxies and with radius in a given galaxy
than any other photometric or spectroscopic index of old populations
\cite{oc92,b3fl,dor95}.  It is evidently extraordinarily sensitive to
the characteristics of its parent population and could therefore be a
uniquely delicate population probe.  But the underlying drivers of the
UVX have not been positively identified.  Ohl et al.\ \cite{ohl97}
have compared internal UV color gradients to abundance gradients and
find little correlation.  Metallicity is evidently not the sole
parameter controlling the UVX.  In the case of M32, which shows a
color gradient reversed from other E galaxies \cite{oc92,ohl97}, an age
gradient may be involved.  

\vspace{3pt}

\noindent {\bf Spiral Galaxies:} The first important conclusion
concerning spiral UV morphologies is that, despite early expectations,
the normal dust distribution in spiral disks does not strongly attenuate
UV light, even in the case of edge-on systems such as NGC 4631 (see Fig.~3).
It appears that the heaviest UV extinction is confined to thin layers
and the immediate vicinity of young H II complexes and that the UV light
emerges from thicker star distributions or regions evacuated of dust
by photodestruction or winds \cite{all97,fot88,calz94,calz97}.

Early-type spirals and barred galaxies tend to exhibit the largest
morphological changes with wavelength.  Figure 4 shows the Sb galaxy
M81, whose Hubble type becomes dramatically later in the UV.  
The optical bulge is dominated by cool stars, and it progressively
diminishes at shorter wavelengths.  In the FUV, the cool MS and
RGB have vanished, but the UVX component of
the bulge is still extended about 30$\arcsec$.  Hot OB associations in
the sprial arms increase in brightness in the UV so that the
galaxy looks like a nearly empty ring system; at optical wavelengths,
such structures would be associated with a short-lived, non-equilibrium
situation.  In other cases (e.g.\ M83), cool bars vanish in the UV.
Van den Bergh et al.\ \cite{vdb96} noted that there are very few, if
any, barred galaxies in the Hubble Deep Field.  Some, though not all,
of this effect may be a morphological k-correction.

\begin{figure}[t!] % 
\centerline{\epsfig{file=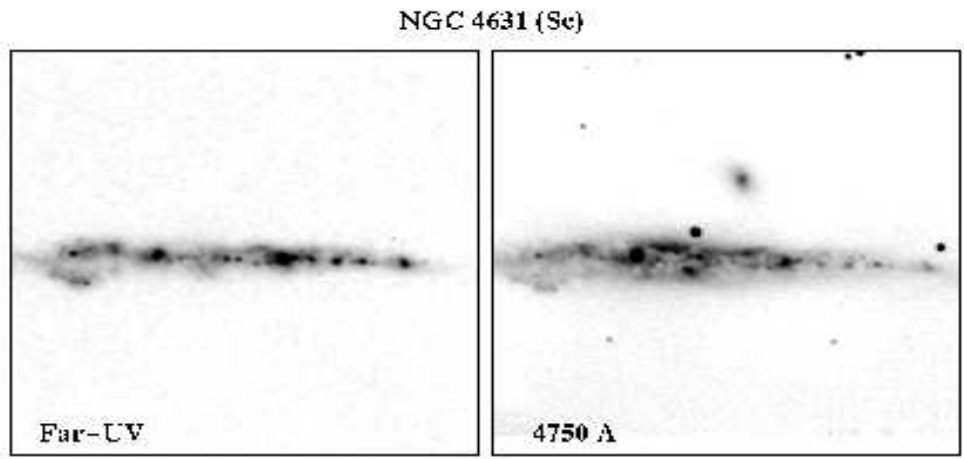,height=3.4in,width=5.1in}}
\vspace*{-0.2in}
\caption{UIT and ground-based images of the edge-on Sc galaxy NGC4631,
exhibiting a classically thin, flat plane of H II regions at both UV
and optical wavelengths.  Extinction by dust is not much more serious
in the UV than in the optical here.  }
\vspace{-0.15in}
\label{fig3}
\end{figure}

\begin{figure}[t!] % 
\centerline{\epsfig{file=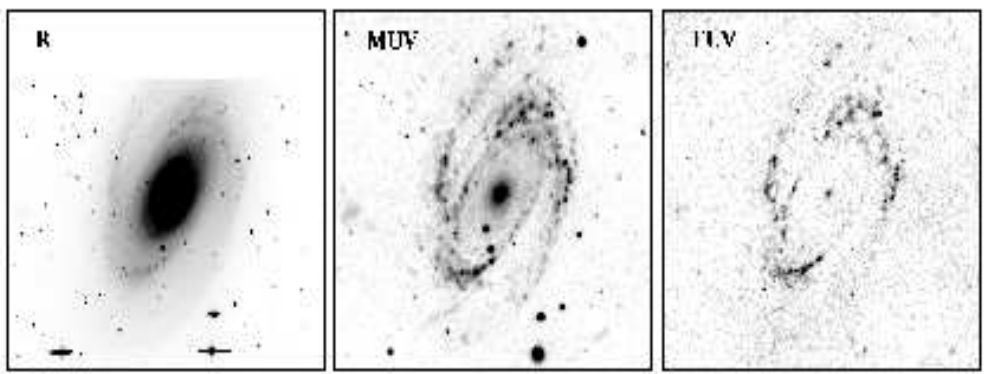,height=2.1in,width=5.2in}}
\caption{UIT and ground-based images of M81.  In the bulge, cool giants dominate
in the R band, the main sequence turnoff dominates
in the MUV, and the UVX produces all the FUV flux.  Hot, massive stars
in the spiral arms become brighter in the UV.  }
\vspace{-0.15in}
\label{fig4}
\end{figure}

The UV light distributions of spirals are, in general, flatter (i.e.\
have longer scale lengths) and more irregular than at optical
wavelengths \cite{lands92,corn94}.  Figure 5 shows the luminous Sc
galaxy M101.  The central concentration strongly diminishes in the UV,
while the prominent OB associations in the disk are emphasized. 
An object like this viewed at high redshift, after surface brightness
and resolution effects are included, gives the appearance of a tidal
interaction between a small swarm of independent galaxies.
Simulations of such effects for M101 and the Sb galaxy M31 are shown
in \cite{boh91} and \cite{om97}.  

Although one normally thinks of massive, hot stars as confined to
compact regions in the spiral arms, in fact, $\ga 75\%$ of the total
far-UV light in M101 is {\it diffuse}.  Only about 25\% of the
H$\alpha$ light is diffuse.  Large far-UV diffuse light fractions have
also been found in M74
\cite{corn94} and M33 \cite{buat94}.  The origin of the diffuse FUV
light---e.g.\ to what extent it may be from grain scattering
as opposed to {\it in situ} hot sources---is not established.

\begin{figure}[t!] % 
\centerline{\epsfig{file=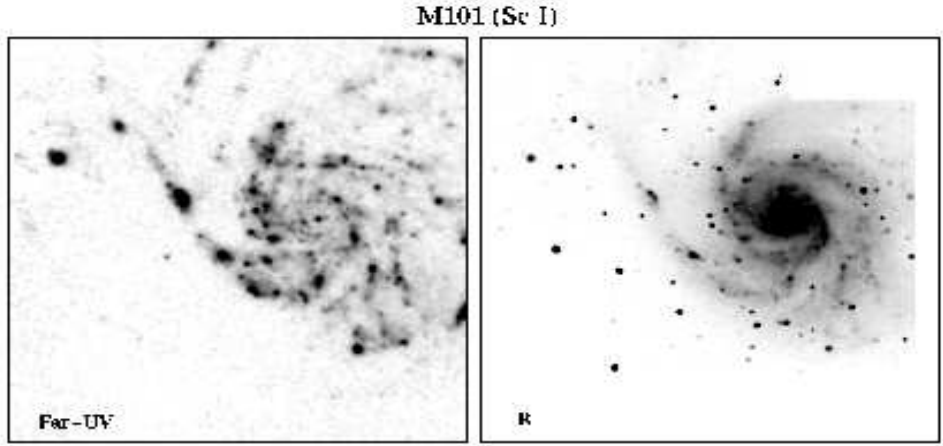,height=3.4in,width=5.0in}}
\vspace*{-0.2in}
\caption{UIT and R-band images of M101.  The brightest OB
association at the upper left is NGC5471.  Significant diffuse far-UV
light is apparent in the disk.}
\vspace{-0.15in}
\label{fig5}
\end{figure}

UV images like these permit us to begin to probe the star formation
{\it history} of disk galaxies on timescales of $\sim 10$--1000 Myr,
intermediate between those relevant to emission lines ($\la 5$ Myr)
and the optical/IR continuum ($\ga 1000$ Myr). Because the luminosity
of a single generation of stars decays roughly as a power law, the
star formation history contributes to light at a given wavelength as
follows:  $L(\lambda) \sim \int SFR\,(t_0 - t)\, t^{-\beta(\lambda)}\,
dt$, where $t$ is the lookback time in Myr and $t_0$ is the present
epoch.  From synthetic models (e.g.\ [1]) $\beta(\lambda)$ is approximately
4.3, 1.5, and 0.8 in the Lyman continuum (Balmer emission lines),
far-UV continuum, and optical continuum, respectively.  By combining
multiband data, one can begin to place meaningful constraints on
$SFR(t)$ over the past few Gyr.  Cornett et al.\ \cite{corn94} used
UIT images to investigate the radial dependence of $SFR(t)$ over the
disk of the Sc galaxy M74 and found a remarkably organized pattern of
change.  The mechanisms by which disks achieve such strong regulation
of star formation are not, I think, well understood \cite{oc97}.

Comparisons of very recent to intermediate-age star formation can be
made by producing H$\alpha$/far-UV difference images, as in Fig.~6 for
M51.  A similar map with lower resolution was published by the FOCA
group \cite{bers94}. One can see how the $\sim\,$50--100 Myr year old
populations (FUV-bright) are usually spread farther downstream from
the density wave than the $\sim\,$5 Myr-old, H$\alpha$-bright populations. 
There is a sense of multiple FUV wavelets here, with feathery
extensions inclined in pitch angle to the main spiral pattern.

\begin{figure}[t!] % fig 6
\centerline{\epsfig{file=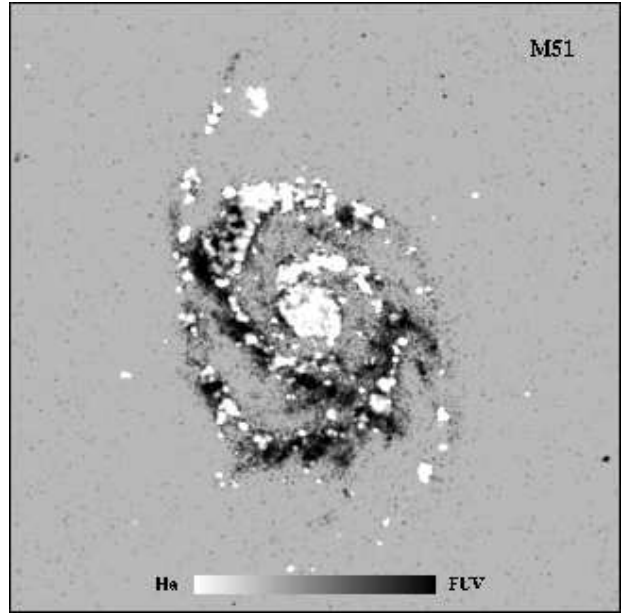,height=3.2in,width=3.2in}}
\vspace{4pt}
\caption{A logarithmic H$\alpha$-far-UV difference map of M51.
Lighter areas are relatively
brighter in H$\alpha$. The map contrasts regions of current star
formation with regions active $\sim$ 50 Myr ago. }
\vspace*{2pt}
\label{fig6}
\end{figure}

\vspace{4pt}

\noindent {\bf Irregular and Peculiar Galaxies:} Except when they have
cool bars (as in the LMC), morphological k-corrections for irregular
galaxies tend to be smaller than for large spirals because even their
optical light is usually dominated by hot stars.  Their UV appearance
can be highly fragmented, however.  The advantages of the UV in
dissecting star formation histories (e.g.\ in the SMC
\cite{corn97}) and detecting very cool or very hot interstellar
material are as described above.   Galaxies in which the normally
thin dust layers have been disturbed by tidal interactions or other
events can have dramatically different UV morphologies.  The best
nearby examples are M82 \cite{courv90,gsh97} and Centaurus A \cite{marc97}.
In M82, the plume of material ejected along the minor axis contains
enough dust to become a reflection nebula at mid-UV wavelengths.   

\vspace{4pt}

With only limited coverage of a small galaxy sample, we have hardly
begun serious UV imaging studies of galaxies.  Deep, all-sky UV surveys
as well as high resolution UV multiband imagery of galaxies out to $z
\sim 0.5$ are badly needed to establish the cosmic history of star
formation and to provide the tools needed to interpret the very distant
universe.  

%%\subsection*{}

%\begin{figure}[t!] % fig 1
%\centerline{\epsfig{file=XXX.eps,height=3.0in,width=3.0in}}
%\caption{    }
%\vspace*{2pt}
%\label{fig1}
%\end{figure}

%%%\section*{Acknowledgements}

\vspace{0.1in}

I am grateful to Pam Marcum, George Becker, Joel Offenberg, and Bob
Cornett for help in assembling the images shown here and to Ted Stecher
and our other colleagues on the UIT team for continuing support.  This
work has been supported in part by NASA grants NAGW-2596 and
NAG5-700.  

\relax

%% ------------------ END TEXT ---------------------------------

%% REFS:  NUMERICAL ORDER

\end{document}